\documentclass[aps,prb,twocolumn,showpacs]{revtex4}
\usepackage{amssymb}
\usepackage{dcolumn}
\usepackage{bm}
\usepackage{amsmath}
\usepackage{amsfonts}

\begin{document}

\title{Comment on: ``Mode repulsion and mode coupling in random
lasers" }
\author{L. I. Deych}
\affiliation{Physics Department, Queens College of City University
of New York, Flushing, NY 11367}

\begin{abstract}
In recent paper Cao et al. [Phys. Rev. B {\bf 67}, 161101 (R)
(2003)] reported an observation of what is the first genuine
multi-mode behavior in random lasers. They observed a splitting of
a single lasing line into two lines with close frequencies when
pumping is increased beyond a certain threshold. Here we are
pointing out that the qualitative interpretation of these
experiments given in that paper is misleading.
\end{abstract}

 \maketitle

In the recent paper Cao, Jiang, Ling, Xu, and Soukoulis (CJLXS)
presented the results of experimental and theoretical studies of
multi-mode behavior in random lasers.\cite{Cao2003} The key
experimental result of that paper was the observation of the
splitting of a single lasing mode into two lines. A frequency
spacing between the new lines was smaller than the homogeneous
line width of the respective optical transitions. Besides it was
found that the temporal behavior of the split-off modes was
synchronized. These experimental findings are the first
demonstration  of a true multi-mode behavior in random lasers, and
are, therefore, of great interest. However, it seems to us that
the theoretical part of Ref.\onlinecite{Cao2003} does not give a
correct physical understanding of these results. The qualitative
interpretation of the experimental results, and subsequent
numerical simulations is given by CJLXS in terms of the mode
coupling. The  CJLXS state that the gain competition, which would
not allow several modes to lase simultaneously, can be overcome by
a coupling between the modes. The nature of this coupling was not
explicitly specified; it was described in rather vague and
non-specific terms of photon hopping and re-emission. If, however,
the term coupling refers to the well-known, in laser physics,
non-linear coupling due to cross-saturation, we would like to
point out that this interpretation is in contradiction with well
established properties of laser oscillations.

According to standard laser theory\cite{Lamb}, the mode
competition arises because of the non-linear mode coupling. This
coupling manifests itself through cross-saturation terms in the
equations of the lasing dynamics. In the simplest case of a
two-mode laser,  neglecting fast oscillating terms, one can
present these equations in the well-known form\cite{Lamb}
\begin{eqnarray}\label{I}
 \nonumber \dot{I_1}&=&2I_1\left(\alpha_1-\beta_1-\theta_{12}I_2\right)\\
    \dot{I_2}&=&2I_2\left(\alpha_2-\beta_2-\theta_{21}I_1\right),
\end{eqnarray}
where $I_{1,2}$ are dimensionless intensities of the respective
modes, and $\alpha_{1,2}$,$\beta_{1,2}$,$\theta_{1,2}$ are linear
net gain, self-saturation, and cross-saturation coefficients
respectively. The latter is responsible for the mode competition
\emph{due to} mode coupling. The simple linear stability analysis
presented in many textbooks on lasing dynamics, for instance, in
Ref.\onlinecite{Lamb}, shows that the dynamics of the modes
depends significantly on the quantity
\begin{equation}
C=\frac{\theta_{12}\theta_{21}}{\beta_1\beta_2}
\end{equation}
called  a coupling coefficient in Ref.\onlinecite{Lamb}. When the
coupling is so strong that $C>1$,  only single mode oscillations
are stable, and one of the intensities $I_{1,2}$ is always equal
to zero. In the case of weak coupling, $C<1$, on the contrary, two
mode solutions are allowed to emerge, when the pumping exceeds
certain threshold value. Thus, the non-linear coupling facilitates
the mode competition rather than quenches it, contrary to what was
assumed in Ref.\onlinecite{Cao2003}. In the light of this
conclusion, the consecutive arguments of CJLXS explaining the
temporal synchronization of the modes does not seem to be relevant
as well.

The simplest, but not the only  possible one, interpretation of
the experimental results of Ref.\onlinecite{Cao2003}, is that
while most of the spatially overlapping modes in random lasers are
strongly coupled, and, therefore, the radiation from all but one
mode is inhibited, the coupling between some modes is weakened
through the mechanism of the spatial hole burning.\cite{Lamb} As a
result, on some rare occasions, two mode oscillations become
possible. In order to see how the spatial hole burning manifests
itself in random lasers, we derived equations similar to
Eq.(\ref{I}) taking into account the inhomogeneous nature of the
background dielectric function $\epsilon(\textbf{r})$, and
assuming that eigen modes, $f_{1,2}(\textbf{r})$, of the passive
dielectric medium are known. The expression for the coupling
coefficient, $C$, was found in this case to have the following
form
\begin{equation}\label{C}
    C\propto\frac{\left[\int \epsilon(\textbf{r})
    f_1^2(\textbf{r})f_2^2(\textbf{r})d\textbf{r}\right]^2}{\int \epsilon(\textbf{r})f_1^4(\textbf{r})d\textbf{r}\int
    \epsilon(\textbf{r})f_2^4(\textbf{r})d\textbf{r}}
\end{equation}
The presence of $\epsilon(\textbf{r})$ in these integrals reflects
a significant difference between regular cavity lasers with a
uniform background dielectric coefficient, and random lasers. If
the modes are well separated spatially, the integral in the
numerator of Eq.(\ref{C}) is small and $C<1$. In this case we have
the typical behavior most frequently observed in random lasers:
many modes lase simultaneously at different parts of the sample
(see Ref.\onlinecite{Cao2003} and references therein). The case
under discussion, however, corresponds to the situation when the
overlap of the modes is rather strong, and the strength of the
coupling depends upon the intricate details of the spatial
distribution of the dielectric function and the eigen mode
functions. Since the two-mode behavior is observed only in very
few cases, the experimental results of Ref.\onlinecite{Cao2003}
indicate that it is much more likely that the modes, which overlap
are coupled strongly, and only for a few rare configurations, the
coupling coefficient drops below unity. In order to verify the
correctness of this assumption careful analysis of statistical
properties of parameter $C$, defined by Eq.(\ref{C}) is required.

This work is partially supported by AFOSR grant F49620-02-1-0305.

\end{document}